# Integration of a 2D Periodic Nanopattern Into Thin Film Polycrystalline Silicon Solar Cells by Nanoimprint Lithography

Islam Abdo[1,2,3*], Christos Trompoukis[1,4], Jan Deckers[1], Valérie Depauw[1], Loic Tous[1], Dries Van Gestel[1], Rafik Guindi[3], Ivan Gordon[1], Ounsi El Daif[5]

[1] IMEC, Kapeldreef 75, 3001 Leuven, Belgium
[2] KACST-Intel Consortium Center of Excellence in Nano-manufacturing Applications (CENA), Riyadh, KSA
[3] Microelectronics System Design department, Nile University, Cairo, Egypt
[4] Katholieke Universiteit Leuven, Leuven, Belgium
[5] Qatar Environment and Energy Research Institute (QEERI), Qatar Foundation, Doha, Qatar

*Corresponding Author, islaam.abdo@gmail.com

## ABSTRACT

The integration of two-dimensional (2D) periodic nanopattern defined by nanoimprint lithography and dry etching into aluminum induced crystallization (AIC) based polycrystalline silicon (Poly-Si) thin film solar cells is investigated experimentally. Compared to the unpatterned cell an increase of 6% in the light absorption has been achieved thanks to the nanopattern which, in turn, increased the short circuit current from 20.6 mA/cm2 to 23.8 mA/cm2. The efficiency, on the other hand, has limitedly increased from 6.4% to 6.7%. We show using the transfer length method (TLM) that the surface topography modification caused by the nanopattern has increased the sheet resistance of the antireflection coating (ARC) layer as well as the contact resistance between the ARC layer and the emitter front contacts. This, in turn, resulted in increased series resistance of the nanopatterned cell which has translated into a decreased fill factor, explaining the limited increase in efficiency.

## I. Introduction

Thin film polycrystalline silicon (Poly-Si) based on aluminum induced crystallization (AIC) and epitaxial thickening could be an interesting photoactive layer for lower-cost crystalline silicon solar cells as it can be grown from vapor phase on non-silicon lower cost foreign substrates. This has been made possible thanks to the AIC layer exchange process [1] producing a 3-micrometer layer of Poly-Si on alumina substrate. AIC is one member of the metal induced crystallization (MIC) family using the aluminum to produce a very thin seed layer which is further thickened by epitaxial growth. The process of fabricating the seed layer is based on the layer exchange between an amorphous silicon (a-Si) thin layer and an aluminum layer deposited on top. By annealing at 500℃ the two layers exchange positions producing a *Coarse-grained* seed layer with grains of 1-100 µm. This two-step approach enables the decoupling of the doping/profile type and the final layer thickness from the initial properties of the formed seed layer like its grain size and preferential orientation [1]. Other alternative one-step techniques were also investigated like thermal solid phase crystallization (SPC) [2], which is the only technology that had already been matured for industrial production [3], laser induced crystallization (LIC) [4], electron beam crystallization [5], joule heating [6]; the techniques which enable the re-crystallization of a deposited Si layer that can directly be used as the base layer in the cell processing [1]. In this work we used the AIC based Poly-Si as our base layer to make use of the advantage of the two-step process approach which had led to an efficiency of 8% obtained by imec, Belgium [7].

Light trapping is an inherent challenge to all crystalline silicon thin-film solar cell technologies due to the poor light absorption of crystalline silicon near its bandgap energy and the typical low thickness of the fabricated base layer. Therefore, an efficient light trapping scheme is inevitable for such thin layers to achieve high solar cell performance. However, there was always a compromise between the optical light trapping efficiency of the used scheme and its material consumption. This tradeoff is obvious in the random pyramid texturing technique which is one of the most extensively used state of the art light trapping techniques for commercial crystalline silicon solar cells. Pyramids of several micrometers high are fabricated using potassium or sodium hydroxide etching [8] on thick crystalline silicon photoactive layers. The optical working principle of this technique is based on the statistical harvesting of light or geometrical optics [9] as the dimension of the pyramids are much larger than the wavelengths range of the solar spectrum. As a result of its high material consumption, this texturing technique cannot be used for ultra-thin film technologies of few micrometers since the material waste during the pyramid formation is in the order of the thin-film thickness. The challenge of reducing material consumption during texturing has partially been solved by the random plasma texturing technique which could create randomly textured surfaces at the nano-scale. However, this technique still consumes ~1 micrometer of material and creates a very rough surface that is difficult to passivate. This texturing technique has been used on the AIC based Poly-Si cells and a current of 20 mAcm⁻², an open circuit voltage of 534mV, a fill factor of 73% and an efficiency of 8% were obtained [7]. Fortunately, progress in nanophotonics made it possible to fabricate periodic nanostructures in the range of wavelengths that silicon absorbs [10], [11]. Many low material consumption techniques for light trapping have been emerged like hole mask colloidal lithography (HCL) [12] and laser holographic lithography [13], [14]. These techniques open the possibility of surpassing the Yablonovitch limit [15] which states that the light path length enhancement obtained by statistical light harvesting techniques is only limited to $4n^2$, where n is the refractive index of the base material. Nanoimprint lithography (NIL) [16] is one of these promising technologies that enable the fabrication of surface nanopattern of sizes in the wavelength range on large area. NIL is a low cost, high throughput, high resolution lithography technique and has already been used to fabricate surface nanopattern for light trapping in thin film crystalline silicon solar cells [17], [18].

In this work, we propose using NIL combined with dry etching to fabricate a two dimensional (2D) periodic surface nanopattern in order to improve the optical properties, and hence more light generated current, of an AIC based Poly-Si solar cell with a photoactive layer of a 3-micrometer thickness. In the next section, the nanopattern and solar cell fabrication details are presented, together with the transfer length method (TLM) that is introduced to characterize the contact and sheet resistances of the front layers of the Poly-Si cells studying the influence of integrating the nanopattern. Then the results of integrating the nanopattern are presented and discussed topographically, optically and electrically.

## II. Experimental Details

As shown in Fig. 1, and also described in [17], the 2D periodic nanopattern is defined in three steps: soft stamp fabrication, NIL using the fabricated soft stamp and finally reactive ion etching (RIE) of the silicon layer. The process starts by fabricating a silicon master stamp patterned by Deep Ultraviolet (DUV) lithography and dry etching to the desired patterning shape and dimensions. A polydimethylsiloxane (PDMS, Sylgard®184 Silicone Elastomer from DOW CORNING) soft stamp is then fabricated using the master stamp as a mold where the soft stamp carries the negative pattern of the master stamps. The Poly-Si samples to be nanopatterned are spin coated with a thermoplastic resist which is then baked at 100℃ to evaporate its solvents. The soft stamp is used to transfer the nanopattern to the resist by using a home-made hydraulic press, in which the resist is deformed by compression. This imprint step is done at 130°C which is above the glass transition temperature of the resist. At that temperature, both the resist Young's modulus and viscosity drop by several orders of magnitude for easy deformation [10]. A two-step process of RIE is then applied to, firstly, etch the residual thin layer of the resist mask using O₂ plasma, and, secondly, a combination of $SF_6$ and $O_2$ to chemically and physically etch the Poly-Si through the nanopatterned resist. The resulting etching profile is totally independent of the crystal orientations of the Poly-Si grains because of the violent ionic bombardment involved in the physical etching part of the process. The dimensions of the nanopatterning can be controlled by the various etching parameters (power, pressure, gas concentration and time). Although the period of the nanopattern is predefined from the used master stamp, its depth and diameter-to-period ratio are controlled by the etching conditions. Finally the resist is removed by dissolution in acetone.

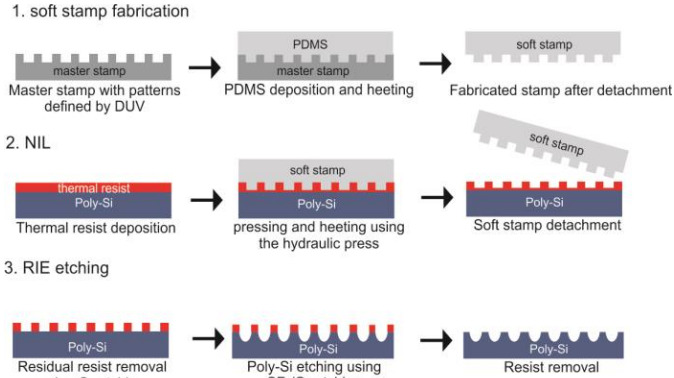

Fig. 1. Soft thermal nanoimprint lithography and dry etching process flow

To fabricate an AIC based Poly-Si solar cell, a heterojunction emitter structure is used [7], [19]. As shown in Fig. 2, the process starts by defect passivation by plasma hydrogenation at 600℃ and then deposition of an intrinsic (10nm) and n+-doped (10nm) hydrogenated amorphous silicon (i/n+ a-Si:H) layer by plasma enhanced chemical vapor deposition (PECVD) for surface passivation and emitter formation, respectively. An 80nm layer of indium tin oxide (ITO) is then sputtered as a transparent conductive oxide (TCO) [20] also acting as an antireflection coating (ARC). The ITO layer material is optimized to allow for high optical transmission as an ARC and high electrical conduction as a TCO. The refractive index of the ITO layer is 2.1 at 550 nm. It stays roughly constant till the bandgap wavelength of Silicon (~1100 nm). The contacts are then formed on the front side of the Poly-Si sample using an interdigitated scheme. The aluminum base contacts are formed by a lift off process, in which the contacts are defined by photolithography. During the photolithography process, two stacked layers of photoresist were deposited in such a way (see step 3 of Fig. 2) that enables the under etching of the ITO layer to avoid its contact with the Al base contacts to be deposited afterwards. The ITO layer and Poly-Si are then chemically etched to allow for the electron beam evaporation of aluminum fingers to contact the base. Ti/Pd/Ag contacts are e-beam evaporated through shadow masks to contact the emitter. Finally, the fabricated cells are annealed at 200℃ after which the cell separation is done by dicing;

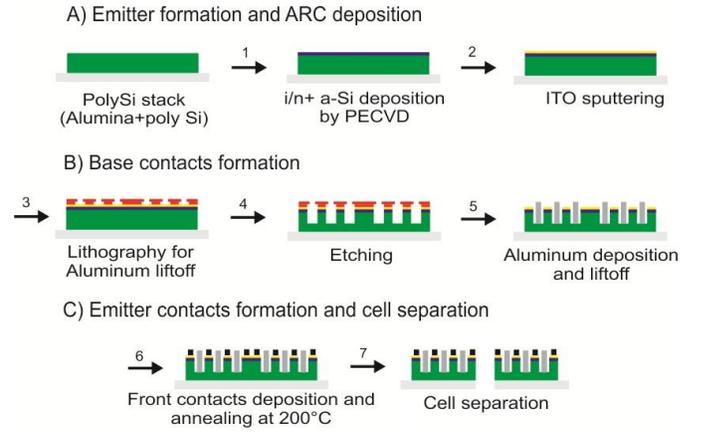

Fig. 2. Poly-Si interdigitated solar cell processing.

The nanopatterning is done just before the i/n+ a-Si deposition for the etched surfaces to be passivated by the hydrogenated intrinsic a-Si which is inevitable after the damage the RIE introduces to the surface. The resulting topography is characterized using scanning electron microscopy (SEM). Optical characterization is done by spectrally resolved reflectance and transmittance using an integrating sphere. The integrated values $(X_{int})$ are calculated over the wavelength range from 300 nm $(\lambda_{min})$ to 1100 nm $(\lambda_{max})$ using the following formula:

$$X_{int} = \frac{\int_{\lambda_{min}}^{\lambda_{max}} \frac{\lambda}{hc} S_{AM1.5G}(\lambda)\, X(\lambda)\, d\lambda}{\int_{\lambda_{min}}^{\lambda_{max}} \frac{\lambda}{hc} S_{AM1.5G}(\lambda)\, d\lambda} \quad (1)$$

Where $X(\lambda)$ is the reflectance, transmittance or absorbance at wavelength $\lambda$, $h$ is Blank's constant, $c$ is the speed of light in vacuum, and $S_{AM1.5G}(\lambda)$ is the air mass 1.5 global tilt solar intensity at wavelength $\lambda$.

The illuminated J-V characteristics of the fabricated cells are measured under AM1.5G spectrum in a calibrated AAA solar simulator. Contact resistance characterization is done using TLM [21].

The TLM is a classical method, proposed by Shockley [22], used to measure the contact resistance $(R_c)$ of a metal-semiconductor interface, but the measurement is independent of the type of contact. The TLM test structure [see Fig. 3(a)] consists of unequally spaced contacts where the contacts have a width W, length L and the separation distance between the fingers is denoted by d1, d2, d3,..di. By the application of a voltage difference on a pair of contacts of separation distance $d_i$, current flows between the two fingers and then the resistance is then calculated, consisting of the specific contact resistance $\rho_c$ between the two layers and the sheet resistance $R_{sh}$ of the bottom layer. The typical curve resulting from the TLM measurements is a straight line whose slope is proportional to $R_{sh}$ and whose intersection with the resistance axis is twice the contact resistance between the two layers. $R_c$, $R_{sh}$ and $\rho_c$ can all be related by the following equation:

$$R_c = \frac{\rho_c}{A_{eff}} = \frac{\rho_c}{L_T W}; \quad L_T = \sqrt{\rho_c/R_{sh}} \quad (2)$$

Where $L_T$ is the transfer length [21], [23], [24].

In the next section we use the TLM method to quantify the impact of integrating the nanopattern into the Poly-Si cell in which measurements are done for the ITO layer and Ti/Pd/Ag emitter contacts since they are the most affected layers by the presence of the nanopattern. Consequently, our TLM structure [see Fig. 3(b)] consists of an ITO sputtered blanket layer and then unequally spaced Ti/Pd/Ag contacts are deposited by electron beam evaporation. The supporting structure is a non-conductive substrate consisting of a 400-um-lowly-doped p-type crystalline silicon wafer coated with a thin 20 nm layer of intrinsic amorphous silicon. This stack guarantees no leakage current flowing into the substrate which could affect the accuracy of the measurements.

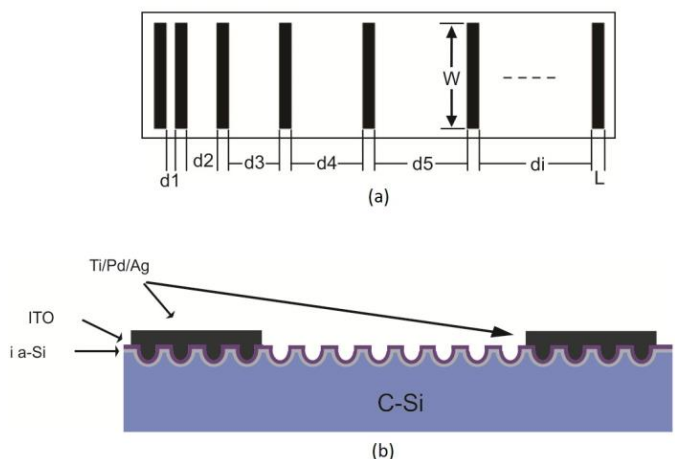

Fig. 3. (a) TLM test structure with unequally spaced contacts, (b) fabricated TLM structure showing two Ti/Pd/Ag contacts over an ITO layer, insulated from the carrier silicon wafer by an intrinsic a-Si layer.

## III. EXPERIMENTAL RESULTS

Fig. 4(a) and Fig. 4(b) show cross section and top view SEM images of a nanopatterned bare Poly-Si layer, respectively. The fabricated nanopattern has a period of 900 nm, a diameter of 800 nm and a depth of 550 nm. The period was predefined from the master stamp while the depth and the diameter were chosen to maintain good optical properties of the nanopattern, giving as high absorption as possible, while keeping in mind that the resulting pattern structure should enable conformal deposition of the intrinsic a-Si:H layer for good surface passivation and good electrical properties. Despite the high roughness of the Poly-Si layer were able to nanopattern more than 70% of the surface [see Fig. 4(b)] thanks to the flexibility of the soft stamp.

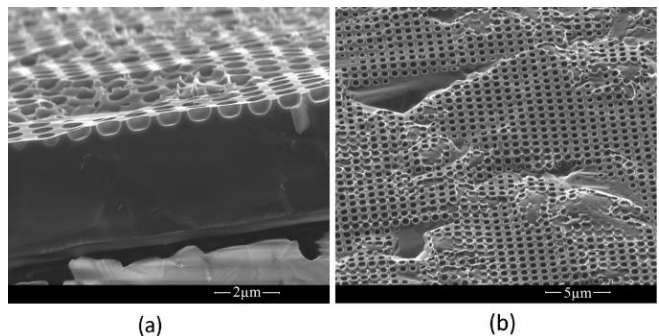

Fig. 4 Topography characterization (a) cross section SEM of a nanopatterned Poly-Si (b) top view SEM where ~70% of the surface area is nanopatterned.

Optically, Fig. 5 shows the reflectance and transmittance of the fabricated Poly-Si cells, with and without nanopatterning. We were able to decrease the integrated reflectance down to 22.4% compared to the unpatterned cell of 27.1% integrated reflectance whereas there is no significant change in transmittance [see Fig. 5 (a)]. Fig. 5(b) shows the absorbance (A) of both types of cells, deduced from the measured reflectance (R) and transmittance (T) through the relation A=1-R-T, where the integrated absorbance, consequently, increases from 68.7% to 74.1%. We notice from Fig. 5 that there is hardly a difference in reflectance, and absorption, at the wavelength range around 500 nm. This can be explained by the fact that the ITO antireflection layer is optimized for a wavelength of 550 nm, and therefore the antireflection effect of the ITO layer is taking place around 500nm, making the contribution of the nanopattern negligible at that wavelength range. In [25] the nanopattern was also compared to the plasma texturing on Poly-Si bare layers with an ARC layer, where it showed that the integrated reflectance of the nanopatterned Poly-Si layer is less by 2% compared to the plasma textured layer.

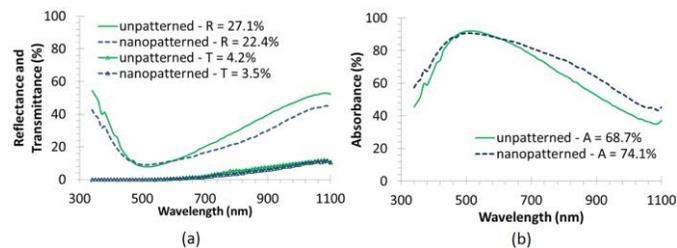

Fig. 5 Optical properties of unpatterned and nanopatterned cells: (a) reflectance and transmittance measurements (b) deduced absorbance spectrum.

The optical characteristics of the unpatterned and patterned cells can be explained by studying short and long wavelengths separately. At short wavelengths, from 300 nm to 650 nm, the reflectance decreased because the nanopattern decreases the amount of the planar part of silicon at the top of the structure to the sub-wavelength range and creates a smooth transition between air and silicon leading to better light coupling inside the photoactive layer as a result of impedance matching between the two mediums [26]. At long wavelengths, on the other hand, light absorption properties is affected by the surface morphology of both the front and the back sides of the cell structure. At the front side, light is diffracted at high angles due to the fact that the pattern dimensions are of the order of the wavelength of incident light, enhancing the interference between transmitted light from the front side and the reflected light from the back side leading to total internal reflections (TIR) assuming a flat back side. However, at the back side we have a rough alumina substrate that can be considered as a Lambertian diffuser. As described in [27], the presence of, *both*, a top grating and a back Lambertian diffuser decreases the TIR inside the structure allowing for high escape transmission through the front side, decreasing the amount of light absorbed in the base layer. This effect can be clearly noticed at the wavelength range around 500 nm where there is no significant difference in reflectance and absorbance. However, we still have enhancement of the total absorption, in spite of these negative effects resulting from the combination of the nanopattern and the back diffuser.

It should be noted that part of the measured light absorption is due to light absorbed in layers other than the photoactive layer [28]. Light at short wavelengths could be absorbed by the ITO and a-Si. Whereas at long wavelengths, it could be absorbed in the alumina substrate and the highly-doped seed layer. The shadowing effect of the metal front contacts also plays a role in decreasing the absorption and decreasing the effective area of the cell.

Electrically, Fig. 6(a) compares the J-V characteristics of the unpatterned and nanopatterned cells showing an increase in the short circuit current ($J_{sc}$), confirming what is expected from the improvements in light absorption. As also shown in Table. 1, the $J_{sc}$ has increased from 20.6 mA/cm$^2$ to 23.8 mA/cm$^2$ which is higher than the $J_{sc}$ obtained in [7] by plasma texturing. This increase in current can be confirmed by the increase in the external quantum efficiency (EQE) shown in Fig. 6(b) due to the nanopattern. EQE has increased at short and long wavelengths, confirming better light coupling into the base layer. However, we can notice that the relative increase in $J_{sc}$ (15.5%) is much higher than the relative increase in absorbance (7.9%). This can be confirmed by the EQE measurements showing an increase at the wavelength range around 500 nm while there is hardly a difference in absorbance, as mentioned before, on the same wavelength range. This suggests that, thanks to the nanopattern, the parasitic absorption in the ITO and a-Si layers is decreased. This might be due to the fact that the ITO and a-Si layers are not conformally deposited on the nanopattern as shown by V. Depauw et al. [28]. This reduces the a-Si thickness by 40% at the side walls of the nanopattern whereas the thickness of the ITO is only 10% of the horizontal thickness. With the fact that we have the same absorbance for both types of cells at that wavelength range and, simultaneously, having lower parasitic absorption in the nanopatterned cell, one might say that this increases the useful absorption in the photoactive layer of the nanopatterned cell generating more carriers that lead to higher current. We notice that there is a reduction in the open circuit voltage ($V_{oc}$) after nanopatterning, but the drop in the $V_{oc}$ is only ~18 mV which is not highly significant for this type of cell, with material nonuniformities. A stronger drop was observed in [28] with a monocrystalline thin film. This limited impact is likely due to the fact that the minority-carrier diffusion length in Poly-Si is already low due to the presence of defects at the grain boundaries as well as intragrain defects [29], which result in a domination from bulk recombination over surface recombination, contrarily to the case of monocrystalline silicon films. That drop might be due to the geometrical increase of the surface area and the surface damage due to the RIE during nanopatterning. However, the contribution of the surface damage due to the RIE is much higher than the geometrical increase of the surface area. This can be confirmed by the minority carrier life time measurements done by C. Trompoukis et al. [30] on nanopatterned float zone (FZ) and Czochralski (Cz) wafers where it was shown that RIE nanopattern significantly decreased the minority carrier life time for both types of wafers. The fill factor (FF) has also decreased from 66.1% to 61.9%, but overall, the conversion efficiency (η) has still increased from 6.4% to 6.7%.

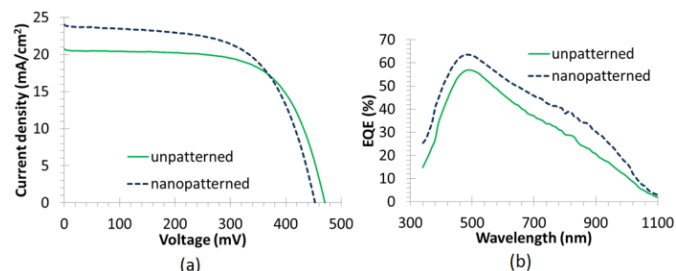

Fig. 6 Electrical performance of unpatterned and patterned Poly-Si cells: (a) J-V characteristics showing an improvement in $J_{sc}$, (b) Improvement in EQE.

| Texturing | $J_{sc} (mA/cm^2)$ | $V_{oc} (mV)$ | $FF$ (%) | $\eta$ (%) |
|---|---|---|---|---|
| Unpatterned | 20.6 | 470 | 66 | 6.4 |
| Nanopatterned | 23.8 | 453 | 62 | 6.7 |

The limited increase in the efficiency is mainly due to the decrease in both the *fill factor* and $V_{oc}$ of the cell. In order to study the effect of the nanopattern on the electrical performance of the cells, the series resistance of the cell with and without the nanopattern was measured. By measuring the J-V characteristics of the cell at different illumination intensities, the technique which was first proposed by Swanson [31], one could measure the series resistance of the cell from the difference in the voltage drop on it at different intensities. Using this technique, we found that the series resistance has increased from 3 Ω to 3.4 Ω justifying the drop in the fill factor.

To further interpret the increase in the series resistance of the nanopatterned Poly-Si cell, a TLM structure is fabricated as described in the previous section to check the impact of the nanopattern on the front layers (ITO and Ti/Pd/Ag). The results [see Fig. 7(a)] show two straight lines corresponding to the unpatterned and nanopatterned cells. The slope of the plotted line is higher for the nanopatterned cell, which is an indication for the increase of the sheet resistance of the ITO layer compared to the unpatterned cell. $R_{sh}$ has increased from 48 Ω/square to 79 Ω/square [see Table. 2]. This increase in the $R_{sh}$ of ITO is due to its poor conformality on top of the nanopattern as was shown in [28] where the nanopattern was integrated in thin crystalline silicon solar cells. The contact resistance between the two layers also increased (from 1.2 Ω to 1.6 Ω) due to nonconformality of both ITO layer and Ti/Pd/Ag contacts, resulting in the formation of trapped air voids in the later [see Fig. 7(b)].

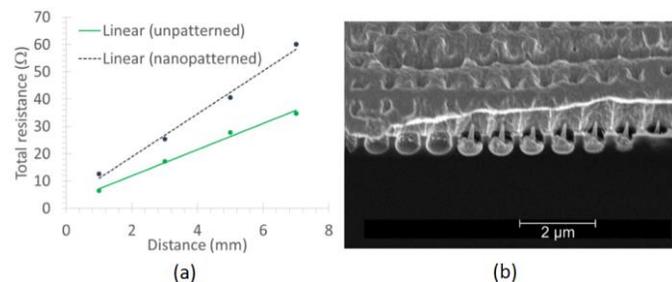

Fig. 7  (a) TLM measurements for Ag contacts over ITO layer showing increase in both $R_{sh}$ and $R_c$, (b) SEM image showing the formation of air voids in the Ti/Pd/Ag contact deposited on top of the ITO layer, both deposited on a nonconductive substrate

TABLE II
TLM MEASURED PARAMETERS FOR TI/PD/AG DEPOSITED ON ITO

| Texturing | $R_{sh}(\Omega/\text{square})$ | $R_c(\Omega)$ |
|---|---|---|
| Unpatterned | 48 | 1.2 |
| Nanopatterned | 79 | 1.6 |

## IV. CONCLUSION

A 2D periodic nanopattern using NIL has been integrated into AIC-based Poly-Si cells boosting the short circuit current. $J_{sc}$ has increased from 20.6 mA/cm² to 23.8 mA/cm² as a result of the enhanced absorbance, beating the $J_{sc}$ value of the record cells textured by plasma thanks to the nanopattern. The fill factor, however, has decreased as a result of the nanopatterning which was shown to increase the sheet resistance of the ITO layer. The increase in the sheet resistance of ITO and the formation of trapped air voids, as confirmed by the TLM measurements and SEM, increased the contact resistance between the ITO layer and the Ti/Pd/Ag contacts. The efficiency of the cell, consequently, has slightly increased from 6.4% to 6.7%.

For efficiency improvements from the electrical side, the $V_{oc}$ could be improved by using Tetramethylammonium hydroxide (TMAH) wet etching  instead of the RIE etching used after NIL [30]. For the FF, using thermal evaporation or sputtering as the deposition technique for the front metal contacts is expected to have a smaller grain size of silver, which could prevent the formation of trapped voids and thus lead to a better contact to the ITO layer. The ITO layer conformality on the nanopattern could be improved by tilting the sample at different angles during the sputtering process using the oblique angle deposition [32]. On the optical side, optimizing the nanopattern dimensions for better light absorption and light coupling into the photoactive layer by the help of the optical simulations done in [26], could contribute to higher $J_{sc}$ value. A possibility would be to use HCL [12] for patterning which was shown in [33] to increase the nanopatterned area and increase it above 70% coverage which has been achieved by NIL. NIL surface texturing could also be integrated into thin film polycrystalline silicon solar cells based on liquid phase crystallization (LPC), the technique that has achieved the largest grain sizes and the best material quality so far. Furthermore, LPC is not only limited to planar surfaces but it also enables the deposition of Polycrystalline silicon on NIL textured glass substrates [3]. This, therefore, enables the combination of both front and back texturing that could result in high efficiency polycrystalline silicon solar cells.

## REFERENCES

[1] D. Van Gestel, I. Gordon, and J. Poortmans, "Aluminum-induced crystallization for thin-film polycrystalline silicon solar cells: achievements and perspective," *Sol. Energy Mater. Sol. Cells*, vol. 119, pp. 261–270, Dec. 2013.

[2] C. Spinella, S. Lombardo, and F. Priolo, "Crystal grain nucleation in amorphous silicon," *J. Appl. Phys.*, vol. 84, no. 10, pp. 5383–5414, Nov. 1998.

[3] C. Becker, D. Amkreutz, T. Sontheimer, V. Preidel, D. Lockau, J. Haschke, L. Jogschies, C. Klimm, J. J. Merkel, P. Plocica, S. Steffens, and B. Rech, "Polycrystalline silicon thin-film solar cells: Status and perspectives," *Sol. Energy Mater. Sol. Cells*, vol. 119, pp. 112–123, Dec. 2013.

[4] P. C. van der Wilt, B. A. Turk, A. B. Limanov, A. M. Chitu, and J. S. Im, "A hybrid approach for obtaining orientation-controlled single-crystal Si regions on glass substrates," in *Proc. SPIE*, San Jose, CA, 2006, p. 61060B-1–61060B-15.

[5] D. Amkreutz, J. Müller, M. Schmidt, T. Hänel, and T. F. Schulze, "Electron-beam crystallized large grained silicon solar cell on glass substrate," *Prog. Photovoltaics Res. Appl.*, vol. 19, no. 8, pp. 937–945, Dec. 2011.

[6] W.-E. Hong, J. Chung, D. Kim, S. Park, and J.-S. Ro, "Supergrains produced by lateral growth using Joule-heating induced crystallization without artificial control," *Appl. Phys. Lett.*, vol. 96, no. 5, pp. 052105-1–052105-3, Feb. 2010.

[7] I. Gordon, L. Carnel, D. Van Gestel, G. Beaucarne, and J. Poortmans, "8% Efficient thin-film polycrystalline-silicon solar cells based on aluminum- induced crystallization and thermal CVD," *Prog. Photovoltaics Res. Appl.*, vol. 15, no. 7, pp. 575–586, Nov. 2007.

[8] K. Birmann, M. Demant, and S. Rein, "Optical Characterization of Random Pyramid Texturization," *26th European PV Solar energy Conference and Exhibition*, 2011.

[9] X. Zhu, L. Wang, and D. Yang, "Investigations of Random Pyramid Texture on the Surface of Single-Crystalline Silicon for Solar Cells," *Proc. in ISES world Congress. 2007 (Vol. I - Vol. V)*, 2009, pp. 1126–1130.

[10] C. Trompoukis, A. Herman, O. El Daif, V. Depauw, D. Van Gestel, K. Van Nieuwenhuysen, I. Gordon, O. Deparis, and J. Poortmans, "Enhanced absorption in thin crystalline silicon films for solar cells by nanoimprint lithography," *Proc. in SPIE*, 2012, Brussels, Belgium, pp. 84380R-1–84380R-10.

[11] J. D. Joannopoulos, Steven G. Johnson, J. N. Winn, R. D. Meade, "Photonic Crystals: Molding the Flow of Light", 2$^{nd}$ ed., Princeton University Press, 2011, pp. 44–121.

[12] H. Fredriksson, Y. Alaverdyan, a. Dmitriev, C. Langhammer, D. S. Sutherland, M. Zäch, and B. Kasemo, "Hole–Mask Colloidal Lithography," *Adv. Mater.*, vol. 19, no. 23, pp. 4297–4302, Dec. 2007.

[13] X. Meng, D. Depauw, G. Gomard, O. El Daif, C. Trompoukis, E. Drouard, A. Fave, F. Dross, I. Gordon, and C. Seassal, "Design and fabrication of photonic crystals in epitaxial-free silicon for ultrathin solar cells," *Proc. in SPIE*, Shanghai, China, 2011, pp. 831207-1–831207-7.

[14] X. Meng, V. Depauw, G. Gomard, O. El Daif, C. Trompoukis, E. Drouard, C. Jamois, A. Fave, F. Dross, I. Gordon, and C. Seassal, "Design, fabrication and optical characterization of photonic crystal assisted thin film monocrystalline-silicon solar cells.," *Opt. Express*, vol. 20, no. S4, pp. A465–A475, Jul. 2012.

[15] E. Yablonovitch and G. D. Cody, "Intensity enhancement in textured optical sheets for solar cells," *IEEE Trans. Electron Devices*, vol. 29, no. 2, pp. 300–305, Feb. 1982.

[16] L. J. Guo, "Nanoimprint Lithography: Methods and Material Requirements," *Adv. Mater.*, vol. 19, no. 4, pp. 495–513, Feb. 2007.

[17] C. Trompoukis, O. El Daif, V. Depauw, I. Gordon, and J. Poortmans, "Photonic assisted light trapping integrated in ultrathin crystalline silicon solar cells by nanoimprint lithography," *Appl. Phys. Lett.*, vol. 101, no. 10, pp. 103901-1–103901-4, Sep. 2012.

[18] E.-C. Wang, S. Mokkapati, T. P. White, T. Soderstrom, S. Varlamov, and K. R. Catchpole, "Light trapping with titanium dioxide diffraction gratings fabricated by nanoimprinting," *Prog. Photovoltaics Res. Appl.*, Oct. 2012.

[19] I. Gordon, K. Van Nieuwenhuysen, L. Carnel, D. Van Gestel, G. Beaucarne, and J. Poortmans, "Development of interdigitated solar cell and module processes for polycrystalline-silicon thin films," *Thin Solid Films*, vol. 511–512, pp. 608–612, Jul. 2006.

[20] A. Stadler, "Transparent conducting oxides—an up-to-date overview," *Materials (Basel).*, vol. 5, no. 12, pp. 661–683, Apr. 2012.

[21] L. K. Mak, C. M. Rogers, and D. C. Northrop, "Specific contact resistance measurements on semiconductors," *J. Phys. E.*, vol. 22, no. 5, pp. 317–321, May 1989.

[22] A. Goetzberger and R. M. Scarlett, "Research and investigation of inverse epitaxial uhf power transistors," Air Force Avion. Lab., Wright-Patterson Air Force Base, O H, 1964.

[23] G. K. Reeves and H. B. Harrison, "Obtaining the specific contact resistance from transmission line model measurements," *IEEE Electron Device Lett.*, vol. 3, no. 5, pp. 111–113, May 1982.

[24] D. K. Schroder, "Semiconductor Material and Device Characterization" 3rd ed., John Wiley & Sons, 2006, pp. 127–149.

[25] I. Abdo, C. Trompoukis, A. Abass, B. Maes, R. Guindi, V. Depauw, D. Van Gestel, I. Gordon, O. El Daif, "Combining periodic nanoimprint lithography and disorder for light trapping in polycrystalline silicon solar cells on foreign substrates," *Proc. in 28th European Photovoltaic Solar Energy Conference and Exhibition*, Paris, France, 2013, pp. 2626–2629.

[26] A. Herman, C. Trompoukis, V. Depauw, O. El Daif, and O. Deparis, "Influence of the pattern shape on the efficiency of front-side periodically patterned ultrathin crystalline silicon solar cells," *J. Appl. Phys.*, vol. 112, no. 11, pp. 113107-1–113107-7, Dec. 2012.

[27] A. Abass, C. Trompoukis, S. Leyre, M. Burgelman, and B. Maes, "Modeling combined coherent and incoherent scattering structures for light trapping in solar cells," *J. Appl. Phys.*, vol. 114, no. 3, pp. 033101-1–033101-11, Jul. 2013.

[28] V. Depauw, X. Meng, O. El Daif, G. Gomard, L. Lalouat, E. Drouard, C. Trompoukis, A. Fave, C. Seassal, and I. Gordon, "Micrometer-Thin Crystalline-Silicon Solar Cells Integrating Numerically Optimized 2-D Photonic Crystals," *IEEE J. Photovoltaics*, vol. 4, no. 1, pp. 215–223, Jan. 2014.

[29] D. Van Gestel and I. Gordon, "Intragrain defects in polycrystalline silicon layers grown by aluminum-induced crystallization and epitaxy for thin-film solar cells," *J. Appl. Phys.*, vol. 105, no. 11, p. 114507-1–114507-11, 2009.

[30] C. Trompoukis, O. El Daif, P. Pratim Sharma, H. Sivaramakrishnan Radhakrishnan, M. Debucquoy, V. Depauw, K. Van Nieuwenhuysen, I. Gordon, R. Mertens, and J. Poortmans, "Passivation of photonic nanostructures for crystalline silicon solar cells," *Prog. Photovoltaics Res. Appl.*, April 2014.

[31] A. Mette, D. Pysch, G. Emanuel, D. Erath, R. Preu, and S. W. Glunz, "Series resistance characterization of industrial silicon solar cells with screen-printed contacts using hotmelt paste," *Prog. Photovoltaics Res. Appl.*, vol. 15, no. 6, pp. 493–505, Sep. 2007.

[32] T. Karabacak and T.-M. Lu, "Enhanced step coverage by oblique angle physical vapor deposition," *J. Appl. Phys.*, vol. 97, no. 12, pp. 124504-1–124504-5, 2005.

[33] C. Trompoukis, I. Abdo, R. Cariou, I. Cosme, W. Chen, A. Dmitriev, E. Drouard, M. Foldyna, E. G.- Caurel, B. Heidari, A. Herman, L. Lalouat, K. Lee, J. Liu, K. Lodewijks, I. Massiot, A. Mayer, V. Mijkovic, J. Muller, R. Orobtchouk, G. Poulain, P. P. Homme, P. Roca, C. Seassal, and J. Poortmans, "Photonic nanostructures for advanced light trapping in thin crystalline silicon solar cells.", Phys. Status Solidi A, Appl. Mater. Sci., Jun 2014.

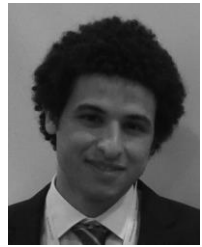

**Islam Abdo** was born in Cairo, Egypt, in 1988. He received the B.S. degree in electronics and communication engineering from Ain Shams University, Cairo, in 2011. He started his MSc. degree in microelectronics systems design in the Nile University, Cairo in 2012 from which he got a 12-month internship to conduct his master's thesis in silicon photovoltaics at imec, Belgium.

From 2012 to 2013, he was a research assistant with the Nile University, Cairo. He is currently pursuing a 6-month internship at imec, Belgium. His research interests includes thin film Silicon solar cells, their fabrication and characterization, and enhancing their optical properties.

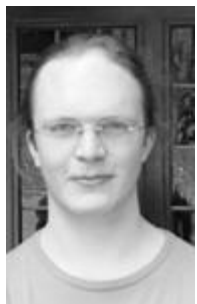

**Jan Deckers** received the bachelor's degree in chemical engineering in 2008 and the master's degree in nanoscience and nanotechnology in 2010. Both degrees were obtained from the University of Leuven (KU Leuven), Belgium.

He is currently pursuing the Ph.D. degree in electrical engineering from the University of Leuven, at imec.

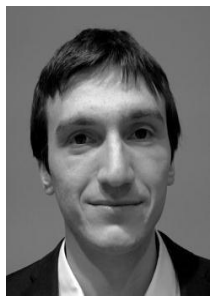

**Loic Tous** was born in Morlaix, France, in 1986. In 2014, he received the Ph.D. degree in Electrical Engineering from KU Leuven, Leuven, Belgium, while he was with imec, Leuven, focusing on nickel/copper plated contacts. He is currently a researcher at imec and his research interests include metallization of silicon solar cells and novel module interconnections.

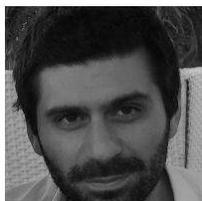

**Ounsi El Daif** is a Physicist. He studied in Lyon, Paris, Beirut, Berlin and Lausanne (Switzerland) where he did his PhD on exciton-photon interactions in semiconductor microcavities. He graduated in 2007. He then went for a post-doc on integrated photonics and nanophotonics for photovoltaics (PV) at the Lyon nanotechnology institute in France. He worked starting 2010 on photonics for solar PV and on thin inorganic PV materials at IMEC in Leuven, Belgium. He is now senior scientist at Qatar Environment and Energy Research Institute (QEERI, Doha, Qatar). Besides research, he teaches Physics at the University of Namur, Belgium and he is interested in bringing a scientific understanding of energy issues to broad audiences, he cofounded therefore the Belgian non-profit organization "L'Heliopole".